# Exploring order parameters and dynamic processes in disordered systems via variational autoencoders

Sergei V. Kalinin,[1] Ondrej Dyck,[1] Stephen Jesse,[1] and Maxim Ziatdinov[2]

[1] Center for Nanophase Materials Sciences, Oak Ridge National Laboratory, Oak Ridge, TN 37831

[2] Computational Sciences and Engineering Division, Oak Ridge National Laboratory, Oak Ridge, TN 37831

We suggest and implement an approach for the bottom-up description of systems undergoing large-scale structural changes and chemical transformations from dynamic atomically resolved imaging data, where only partial or uncertain data on atomic positions are available. This approach is predicated on the synergy of two concepts, the parsimony of physical descriptors and general rotational invariance of non-crystalline solids and is implemented using a rotationally-invariant extension of the variational autoencoder applied to semantically segmented atom-resolved data seeking the most effective reduced representation for the system that still contains the maximum amount of original information. This approach allowed us to explore the dynamic evolution of electron beam induced processes in a silicon-doped graphene system, but it can be also applied for a much broader range of atomic scale and mesoscopic phenomena to introduce the bottom-up order parameters and explore their dynamics with time and in response to external stimuli.

**Introduction**

Over the last two decades, scanning transmission electron microscopy (STEM) has emerged as an indispensable tool for exploring materials structure and functionality on the atomic level.[1-4] Recent advances in aberration correction enabled determining the position of individual atomic columns in 3D materials[5] and single atoms in layered materials[6,7] with picometer-level precision,[8] enabling visualization of ferroelectric polarization and octahedra tilting fields in perovskites,[9-11] symmetry breaking phenomena in complex materials such as ferroelectric relaxors and Kitaev materials,[12] and strain fields in multicomponent systems or in the presence of structural defects.[13] Equally impressive are the advances in the spectroscopic modes such as electron energy loss spectroscopy (EELS), where advances ranging from single-atom spectroscopy[14,15] to mapping plasmonic excitations and phonons have been demonstrated.[14,16-21] Finally, 4D-STEM methods now offer a pathway to further increasing the information limit in STEM,[22-24] providing yet unseen details of atomic-level fields and functionalities.[22,25-27]

The intriguing and yet largely unexplored opportunity presented by recent progress in STEM is the exploration of atomic-scale chemical processes, either induced by classical global stimuli like temperature or induced by electron beam (e-beam) irradiation. Notably, e-beam damage in electron microscopy has been known from the earliest days of the field[28-30] and is traditionally perceived as a strongly deleterious effect. In fact, minimizing e-beam damage along with increasing the spatial resolution, were major factors driving the development of electron microscopy instrumentation. The advent of aberration correction and the potential for atomic-resolution imaging at low voltages it has unlocked, have enabled opportunities toward systematic studies of e-beam-induced reactions. Early examples of these studies include phase transformations in 3D materials[31-34] and e-beam induced radiolysis.[35] A particularly broad set of opportunities has emerged with extensive studies of 2D materials, where nearly all atomic units are available for observation[4] and multiple studies of vacancy and defect formation,[36-41] uncontrolled and controlled atomic motion[42-51] have been reported. Harnessing the electron beam effects on atomic dynamics opened a pathway towards atom by atom manipulation, and recently homo- and heteroatomic cluster assembly have been reported.[52-55]

However, further progress in the field necessitates a quantitative description of e-beam-induced transformations as a necessary step for mapping e-beam-induced reaction networks, comparison with predictive and descriptive theory, and ultimately enabling control of e-beam-induced transformations. In cases where the e-beam-induced changes are localized on a single atom level, such descriptions are possible using the classical point defect chemistry approach, as demonstrated, for example, by Kirkland and co-workers for graphene[36,39] and Maksov *et al.* for layered transition metal dichalcogenides.[56] However, the analysis becomes considerably more complicated in cases where large-scale changes in the connectivity of the chemical bond network occur, including changes in coordination, formation of defect agglomerates, and new phase formation. This is unsurprising since the description of the structure of disordered or partially disordered systems, ranging from amorphous solids to physical systems such as spin and cluster glasses or phase separated oxides, has remained one of the most complex areas of condensed matter physics.[57-59]

In chemically ordered systems, i.e., in cases where the chemical bond network is that of an ideal solid, symmetry breaking phenomena can be effectively analyzed using machine learning tools such as multivariate statistical methods or more complex autoencoder or variational autoencoder based approaches. In this case, symmetry-breaking distortions are referenced to the known (from the global average) ideal lattice, allowing for straightforward analysis. Recently, this approach was applied to the structural description of perovskites based on column[11,60] and unit cell[61] shape analysis, and ferroelectric materials and layered chalcogenides based on the analysis of local atomic neighborhoods.[62] However, this approach fails when the chemical bonding network is non-periodic, precluding the unambiguous definition of the high-symmetry state.

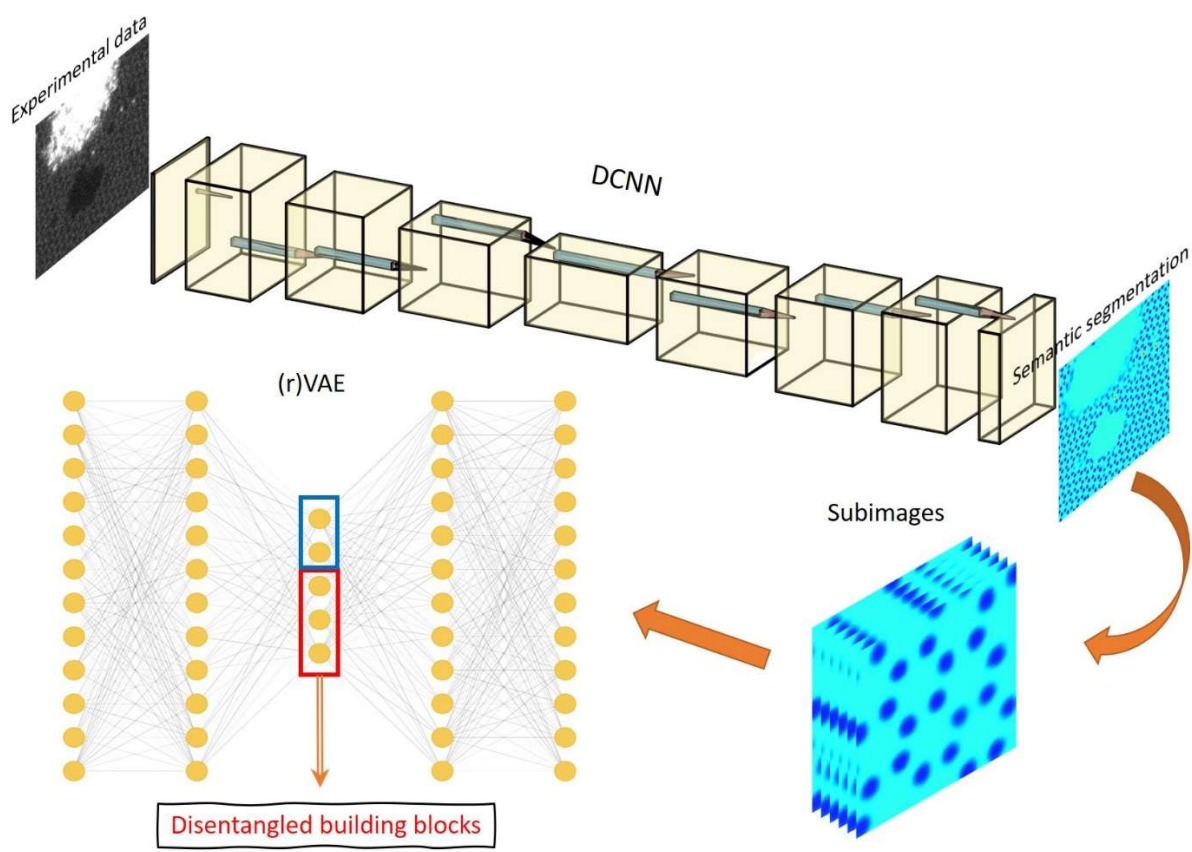

**Figure 1.** Schematic of the overall approach. It starts with passing raw experimental data through a pre-trained deep (fully) convolutional neural network (DCNN) that categorizes every pixel in the data as belonging to a particular atomic type (e.g. C or Si) or to a "background". The DCNN output is a set of well-defined blobs on a uniform background whose centers correspond to atomic coordinates. Next, a stack of sub-images cropped around the identified atomic positions is created and used as an input to a (rotationally-invariant) variational autoencoder ((r)VAE) that seeks to find the most effective reduced representation for the system (building blocks) that still contains the maximum original information. Here, the three neurons in rVAE absorb the information about the rotations and *xy*-translations of sub-image content and the other two neurons are used for disentangling the image content itself.

Here, we explore the applicability of machine learning methods, specifically rotationally invariant variational autoencoders, toward the analysis of chemical transformations in 2D materials under e-beam irradiation. This approach is underpinned by two concepts, the postulated parsimony of structural descriptors in solids and the expectation of global rotational invariance of the system undergoing chemical transformations. That is to say, this approach is rotationally invariant in the sense that we do not know or prescribe the rotation angle of the local bonding pattern with respect to the image plane and allow for it to change from point to point. We demonstrate that the Variational Autoencoder (VAE) with rotational invariance enables the effective exploration of the chemical evolution of the system based on local structural changes and may be extended to more complex systems.

**Results and Discussion**

As a model system, we explore Si in graphene, which has been extensively studied previously in the context of e-beam atomic manipulation.[46,48,49,51,63,64] The description of the experimental protocol is available in the Methods section. The experimental data sets representing the multiple snapshots of the system during atomic manipulation or e-beam-induced decomposition have been analyzed using deep convolutional neural networks (DCNNs) to yield pixel probability maps of the carbon and silicon atoms. Training and tuning of the DCNNs is described in previous publications[56,65-68] and associated codes are freely available as a part of AtomAI package on GitHub.[69]

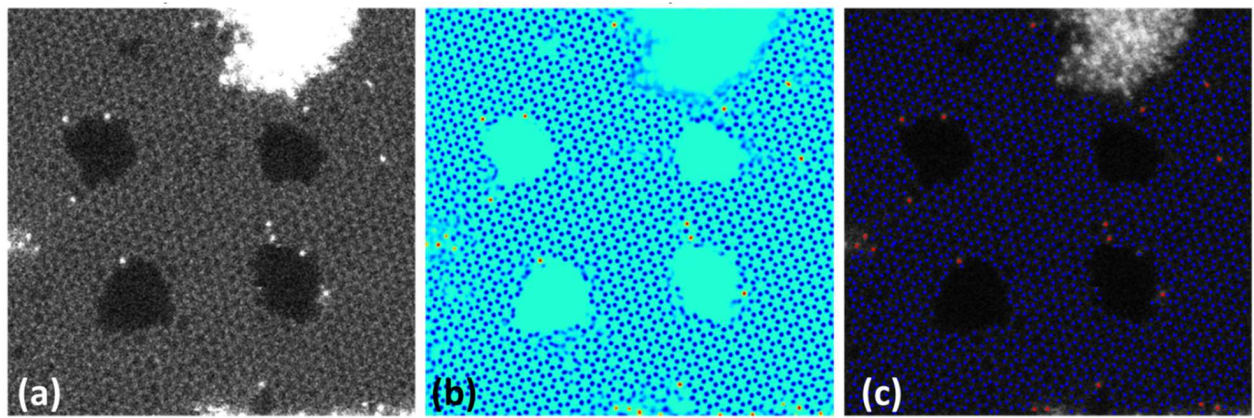

**Figure 2. Experimental and deep learning-processed data.** (a) Single image from dynamic STEM dataset corresponding to 10$^{th}$ frame. (b) DCNN analysis of image in (a) illustrating pixel probability density corresponding to carbon (blue) and silicon (red) atoms. (c) Atomic coordinates of C and Si. Note that amorphous region (same as holes) remain unrecognized by the DCNN. Image size is 16 nm.

An example of the DCNN analysis of dynamic STEM data is shown in Fig. 2. The original STEM image contains clearly visible regions of pristine graphene, regions of agglomerated point defects that may cause rotation and fragmentation with respect to original lattice, multiple holes in the graphene, individual Si atoms (bright dots), and regions of an amorphous Si-containing contamination phase. To analyze the data, we used a fully convolutional CNN with an encoder-decoder type of architecture that takes the raw experimental images and produces images where the values at pixels correspond to their likelihood of corresponding to atoms.[61,70] The DCNN was trained using images simulated by a MultiSlice algorithm[71] and further augmented by applying random cropping, rotations, flipping, scale jittering, adding Poisson/Gaussian noise, and varying the levels of contrast. It is important to note that the network output can be interpreted as a probability that a specific image pixel belongs to a particular atom class, i.e., pixel-wise semantic analysis has been performed.

The DCNN output is shown in Fig. 2 (b). Here, clusters corresponding to carbon (blue) and silicon (red) are visible, which can naturally be interpreted as localization of the corresponding atoms. We define the centroids of the clusters as atomic positions as visualized in Fig. 2 (c). Note that this process is robust and only individual atoms are identified and classified, whereas holes and regions where the graphene is obscured by amorphous contamination are excluded from the analysis. This workflow can be applied to the full dynamic STEM image stack in a fully automated manner, decoding the sequence of frames into time dependent coordinate arrays for C and Si.

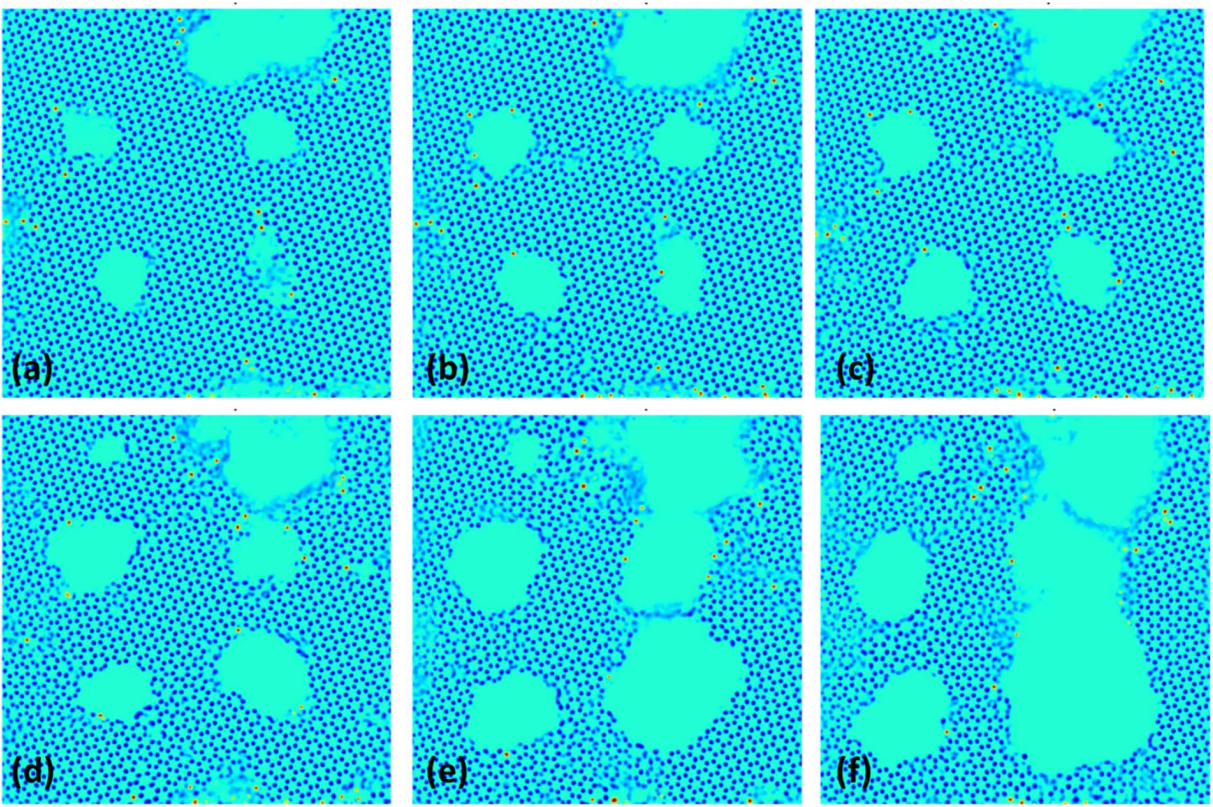

**Figure 3.** Evolution of graphene under e-beam irradiation. Images represent DCNN output at (a) initial, (b) 5$^{th}$, (c) 10$^{th}$, (d) 20$^{th}$, (e) 35$^{th}$, and (f) 49$^{th}$ frame. Red color encodes probability density that specific pixel belongs to Si atom, and blue that it belongs to carbon. Image size is 16 nm.

The evolution of the Si-graphene system throughout the dynamic STEM measurements is shown in Fig. 3. Here, we use the DCNN output similar to that in Fig. 2(b) to illustrate the process. Prior to the measurement sequence, several large-scale holes were created using electron beam manipulation to seed the process. On e-beam irradiation, the formation of 5-7 member defect chains from the introduction, diffusion, and accumulation of carbon vacancies are observed (Fig. 3 (f)).[72] This process is further associated with growth of the holes and formation of small-angle grain boundaries. The Si atoms tend to migrate during the process and segregate to the edges of the graphene sheet (edges of holes). Overall, the process illustrates changes of the connectivity of C-C and C-Si bond networks during e-beam irradiation.

The dynamic changes observed during the e-beam-induced reactions necessitate the development of suitable descriptors, both to quantify the reaction process and distinguish reaction processes in different settings such as temperature and beam fluency, and ultimately develop strategies for guiding the process through global and local stimuli. Here, the use of classical physical descriptors based on lattice symmetry and symmetry breaking phenomena is impossible since the process is associated with the formation and breaking of chemical bonds. At the same time, descriptors based on point defects and clusters in the ideal host lattice also have limited applicability. While individual structural elements such as the formation of 5-7 or 7-5-5-7 defects can be identified, generally the process associated with the formation of complex defect chains and bonding patterns cannot be uniquely represented as point defects.

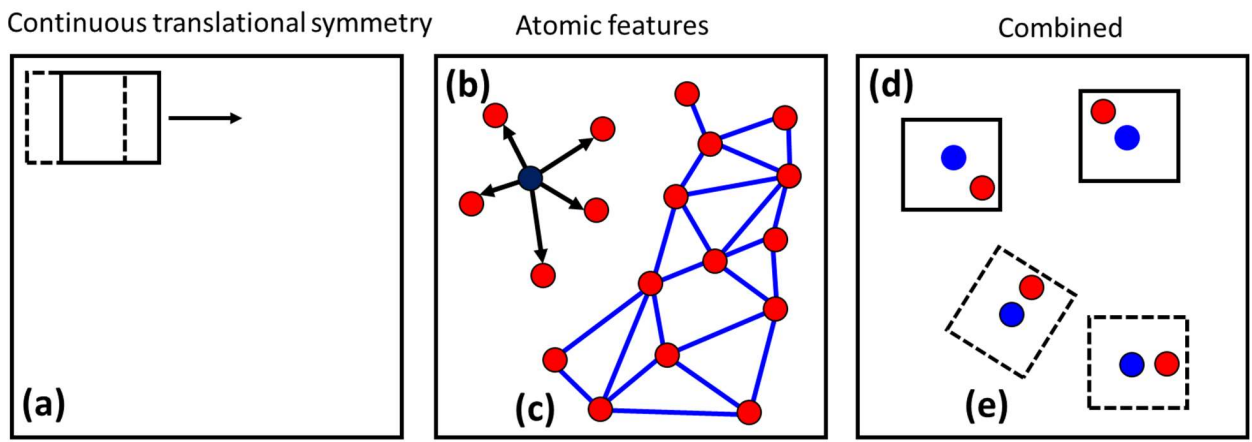

**Figure 4.** Defining the building blocks for machine learning based on extant discrete and continuous symmetries in the system. (a) In systems with continuous translational symmetry, the initial building blocks can be built using convolution with sliding window and subsequent dimensionality reduction using linear or non-linear methods. Note that the same principle is used

in convolutional neural networks. (b) In systems with known atomic positions (or other physically defined reference points), the analysis can be performed either based on the local atomic neighborhood of each atom or (c) analysis of the connectivity graph between point objects. In this approach, the image is reduced to the coordinates of a small number of features and remaining information is discarded. Finally, local descriptors can be built based on (d) continuous neighborhoods of individual atomic units (or other reference point); however, the key element in dimensionality reduction then becomes presence of (e) general rotational symmetry accounting for multiple potentially identical units having different orientations in image plane.

Here, we explore the bottom-up description of the process using a machine learning approach where we do not prescribe *a priori* all of the descriptors; rather descriptors are determined by prominent patterns within the data set. Such an approach can be based on several types of descriptors based on the presence of continuous or discrete translational symmetry and initial local structure representation. For uniform sampling used for continuous translational symmetry as shown in Fig. 4 (a), the sub-images are selected and centered on the rectangular grid in the image plane. Subsequent transforms such as Fast Fourier Transform or Radon transform (depending on the preferred features) combined with linear or non-linear unmixing yields the spatial maps of the descriptive parameters.[73-75] Alternatively, these images can be used as inputs in classical CNNs. This type of analysis is extremely powerful for the detection of patterns such as ferroelectric and magnetic domains, self-assembled monolayers, and atomically resolved images, but generally does not take advantage of specific physically defined features within the data such as individual atom positions. Furthermore, it tends to yield a spectrum of descriptors that are amenable only to semiquantitative interpretation.

An alternative set of approaches can be developed for images where atomic positions (or other discrete reference features) are known, naturally extending to the analysis of sub-images centered on individual atomic coordinates. In a point representation as introduced in a local crystallography approach, a statistical analysis of the radial vectors to the nearest neighbors is performed (Fig. 4b), enabling classification of local structures.[76,77] This analysis can be further extended toward a graphical representation (Fig. 4c). A bonding network can be constructed with the atoms as vertices; the bond strength can be defined as a certain function of atomic spacings and dihedral angles and graph decomposition methods can be employed for deeper analysis. However, this approach relies on the atomic positions being known (or postulated based on a certain threshold level of contrast) and descriptors being the positions of these features. While ideal for theoretical structural data sets, for experimental data the methods that utilize partial or uncertain information on the atomic species positions that allow to work with full images and hyperspectral images and utilizing the statistical representations of such distribution will be highly beneficial.

Here, we explore the approach for identification of the individual elements of the chemical bonding network in the material and its evolution during the reaction based on the combined description, Fig. 4 (d, e), where the sub-images are centered on known atomic positions and the atomic neighborhood is represented as a continuous function defined in the image plane. This approach naturally allows the exploration of images containing both well-defined atomic species and diffuse or uncertain backgrounds. Once the set of sub-images is identified, it naturally brings forth the question of the construction of elementary building blocks or finding a minimal set of descriptors that can represent the full information in the image.

For the system where the chemical bonding pattern does not change, the analysis can be performed using linear dimensionality reduction methods such as principal component analysis or non-negative matrix factorization, Gaussian mixing model (GMM) separation, or non-linear methods such as classical and variational autoencoders as explored earlier for local crystallography and FerroNet[61] approaches. However, the system that has significant orientational disorder will not be amenable to such linear methods since the number of classes is no longer fixed and the decomposition components will be dominated by the rotational degrees of freedom and yield smooth distributions of rotational variants. The example of the GMM analysis is shown in the accompanying notebook. Similarly, analysis through the classic autoencoder yields a complex set of descriptors that convolute both the geometry changes and rotations of individual fragments. This results in different descriptors for structures that only differ by virtue of a rotation (i.e., structures that would likely be considered the same) and separating unique geometric configurations from simple rotations becomes challenging.

To address this challenge, we introduce an approach for analyzing the individual building blocks using rotationally invariant variational autoencoders. The classic autoencoder (AE) operates on the principle of compressing data to a small number of latent variables and subsequently decoding back to the original data set. Compression and decoding are typically performed via a set of convolutional filters similar to classical CNNs. The assumption is that if the information can be represented by a small number of latent variables, then variation of these latent variables across the initial dataset provides insight into the relevant physics. In variational autoencoders (VAE),[78] one assumes that each point in the dataset is generated by a local random variable in a complex, non-linear way. The encoder part of a VAE is used to find good values for the latent random variables such that they are true to our (fixed) prior beliefs and true to the data. The priors in a VAE are the type of distribution for the latent variables (usually a Gaussian distribution) and the parameters of this distribution. The decoder part of a VAE is used to make a prediction for the data based on a sample from this distribution. The model parameters (weights and biases) are learned by maximizing the evidence lower bound (ELBO), which consists of the Kullback-Leibler divergence term and the expected likelihood (reconstruction loss). Note that despite the close similarity in names, AEs and VAEs belong to the fundamentally different classes of machine learning algorithms, with VAEs having the Bayesian nature with the latent space defining the priors for the observables. In the absence of prior knowledge, these are chosen as Gaussian (or uniform); however, more complex models are possible in principle.[79]

Here, the VAE architecture is expanded to account for rotational invariance, which allows rotations to be categorized separately. The implementation of the rotational VAE (rVAE) is based on recent work by Bepler et al.[80] who showed that parameterization of the VAE's decoder as a function of the image coordinates allows separating the image content from rotations and translations. In this case, the latent variable "layer" contains the rotational angle and two translation vectors as three of the latent variables. Additional latent variables are chosen to represent the variability within the (sub-)images content. We found that when using the DCNN output instead of raw data, a simple 2-layer perceptron with 128 neurons in each layer of both encoder and decoder is sufficient to learn the rotationally invariant latent representation of atom resolved STEM data. For the raw data, the rVAE tends to work better with multiple stacked convolutional layers in the encoder. The encoder and decoder networks are trained jointly, using the Adam optimizer for adjusting weights. As suggested in the original work, the rotational prior was modeled using the Gaussian distribution, with zero mean and a standard deviation of 0.1 radians. The rVAE was trained using 75% of the data, while the remaining 25% were used for testing. The rVAE predictions were made on the entire (0.75 + 0.25) dataset.

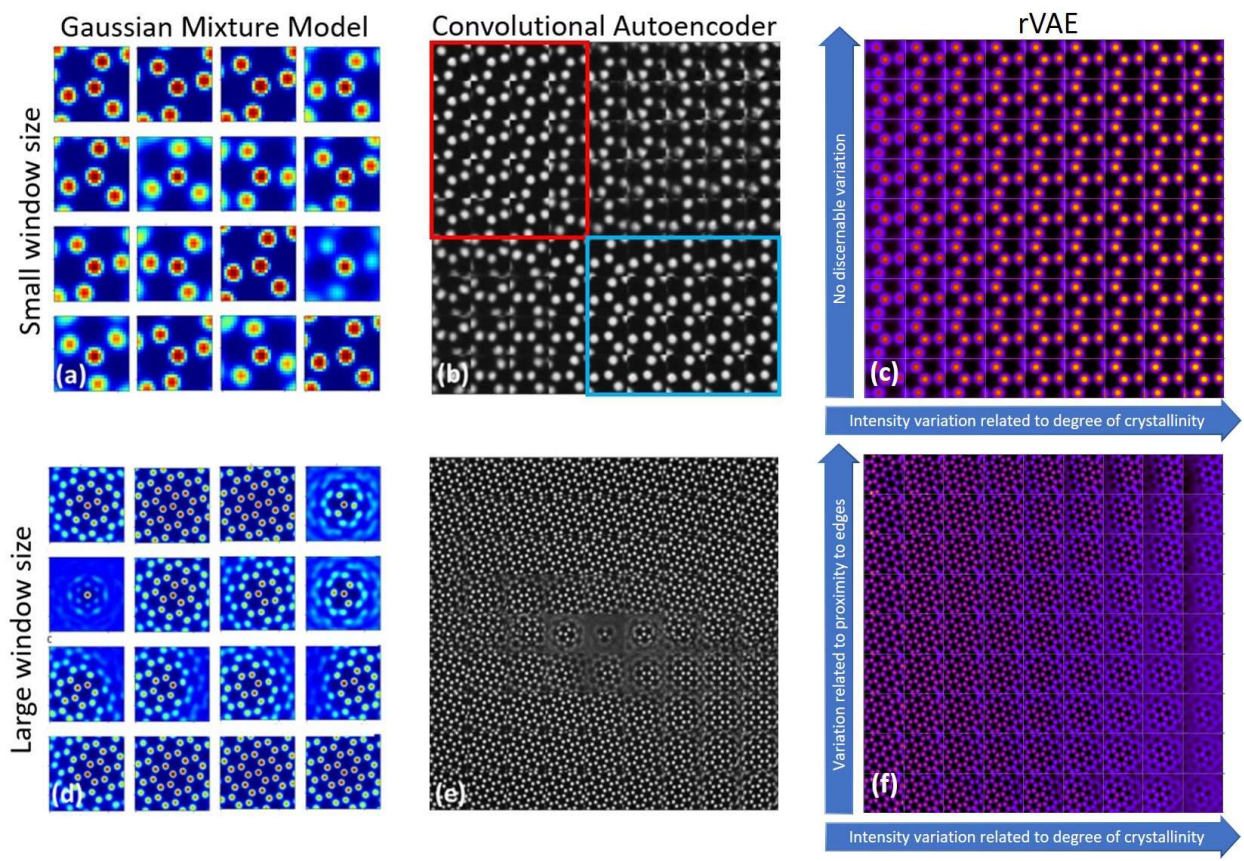

**Figure 5.** Comparison of methods for construction of elementary descriptors for (a,b,c) small and (d,e,f) large windows. (a,d) Gaussian mixture model (GMM) classes of the data that is decomposed into many independent components that are statistical in nature and often does not allow for direct physical interpretation. In (d), this approach performs poorly at capturing rotation and spreads this information across several components. (b, e) Representation in 2D latent space of convolutional autoencoder. Red and blue regions in (b) indicate clear separation of graphene sublattices with remainder of the descriptors encoding lateral shifts, defects, and rotations in a convoluted fashion. In (e), the larger window introduces variability that is more difficult to interpret. (c, f) Representation in the 2D latent space of rotationally invariant variational autoencoder. Since rotational variation is removed from elementary descriptors, remaining variations within data can be described much more efficiently. In (c) there are noticeable changes in only one dimension, which can be ascribed to degree of local crystallinity. In (f), the larger window size also captures variations related to proximity of edges. In (b, c, e, f) the images were generated by applying a corresponding decoder to the uniform grid of discrete point in the latent space.

Figure 5 shows the analysis of the dataset from Fig. 3 using the GMM, AE, and rVAE methods to contrast the effectiveness of each method. The data is a 3D stack of ~80,000 sub-images generated from the 50 initial movie frames. The GMM analysis for a small window, Fig. 5 (a), generates a set of independent components that can be used to linearly reconstruct the best approximation of the image data. Within these components we can see that most correspond to partial rotations of the lattice with some variation in intensity. In comparison, a classical AE analysis is illustrated in Fig. 5 (b) where the full variety of the data contained in sub-images is reduced to two continuous latent variables represented on the $x$ and $y$ axes. In this case, the elementary building blocks can be visualized in the lattice space. Examination of the data shows that the building blocks corresponding to two carbon sublattice sites are localized in the top left and bottom right corners. Intermediate states occupy the diagonal capturing rotation and other more subtle variations. Again, the primary differentiating factor between the building blocks is rotation and the subtle variations do not have a clear physical interpretation. These behaviors become even more pronounced for larger numbers of GMM components or larger window sizes, Fig. 5 (d), which enable a more precise reconstruction of the original data but decrease the physical interpretability. While the GMM effectively determines different components of the data set from an imaging standpoint, these components are mostly the same from a structural standpoint. If the goal of the analysis is to categorize structural evolution of the sample, it is not clear how to proceed using the GMM approach.

It could be argued that using only two latent variables in classical AE may be insufficient to describe the variability of the system. For higher dimensional latent spaces, we adopt the approach where the point distribution corresponding to the real images is clustered via GMM in latent space and the images corresponding to GMM centroids are reconstructed using the decoder part of the AE.[81] We found that generally the reconstructed components also demonstrate the presence of multiple rotation angles. This is unsurprising since changes in the carbon-carbon bond length or geometries are relatively small whereas transformations associated with the rotation between similar structures are large. In other words, rotations do represent the most pronounced feature within the image data even if the structure is unchanged.

The rVAE analysis is illustrated in Fig. 5(c, f). Here, the rotations of structural elements are captured in a separate latent variable allowing for their reliable separation from more subtle structural changes. For smaller window size (Fig. 5c), the variability across the latent space in the $x$ direction is due to intensity, which we associate here with a degree of crystallinity (see the next paragraph), and variation along the $y$ direction is minor. A similar analysis can be performed for much larger window sizes as shown in Fig. 5 (f), demonstrating excellent recovery of structural elements. In this case, one of the latent variables can be interpreted as a degree of crystallinity in the bulk and the other as proximity to the edge of the graphene. The quantitative comparison of GMM, VAE and rVAE using the structural similarity measure between the unmixed/disentangled structures is shown in Supplementary Figure 1.

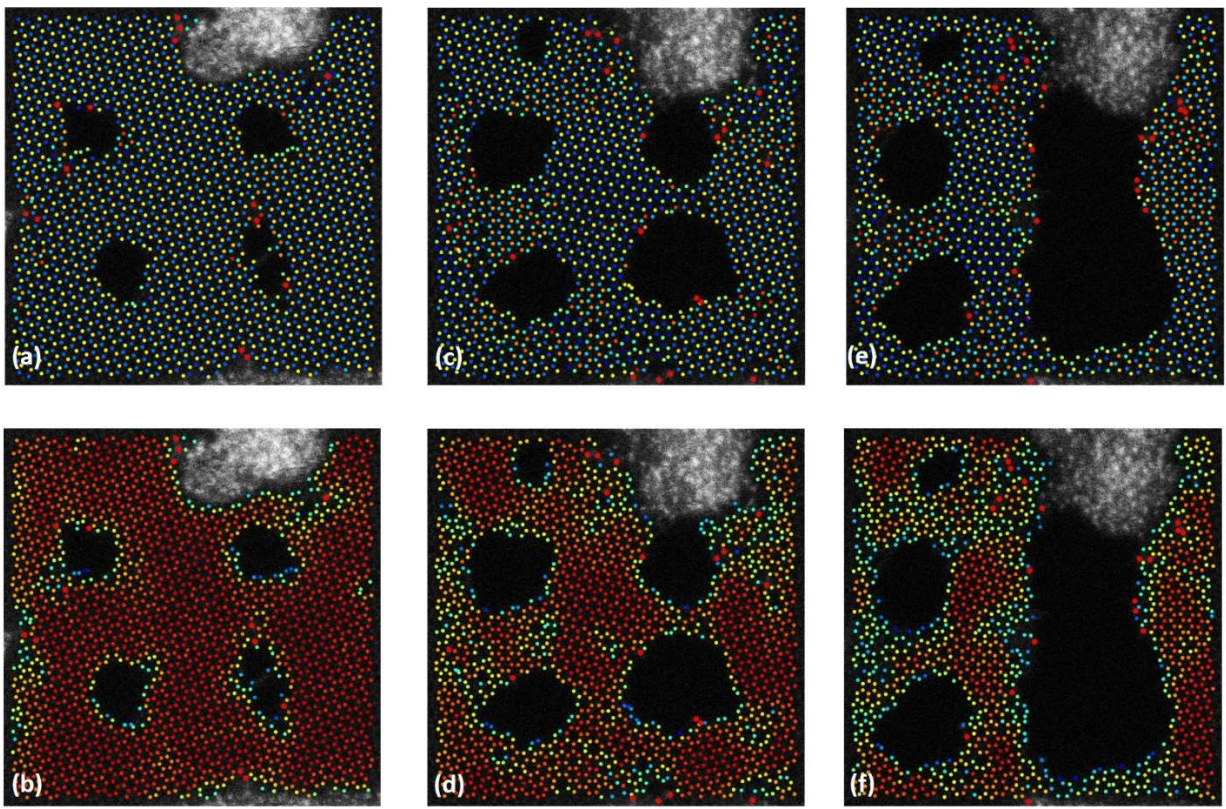

**Figure 6.** Real-space visualization of the rVAE's latent variables. Shown are (a,c,e) angle and (b,d,f) first latent variable for (a,b) initial, (c,d) $25^{th}$, and (e,f) final frame. The positions of Si atoms are indicated by larger red dots. Image size is 16 nm.

Finally, the latent variables can be back projected to real space to visualize the evolution of the atomic structure, as shown in Fig. 6. Here, the color scale of each atomic unit is set according to the value of the corresponding latent variable for angle (top row) and first latent variable associated with images content (bottom row). The angle variable shows clear contrast for the two structurally unique sublattice positions in the graphene lattice, giving rise to the checkerboard-like pattern. Small angle boundaries and the rotation between graphene fragments is clearly visible from the color changes in Fig. 6 (c, e). At the same time, the behavior of the latent variable corresponding to the content of sub-images clearly shows that it adopts maximum value in the well-ordered regions and is reduced at the edges and in regions with high defect density, allowing for its interpretation as local crystallinity. Interestingly, we were able to reproduce these results even when rVAE was trained using only ~25% of the images from each movie frame and then applied to the entire set.

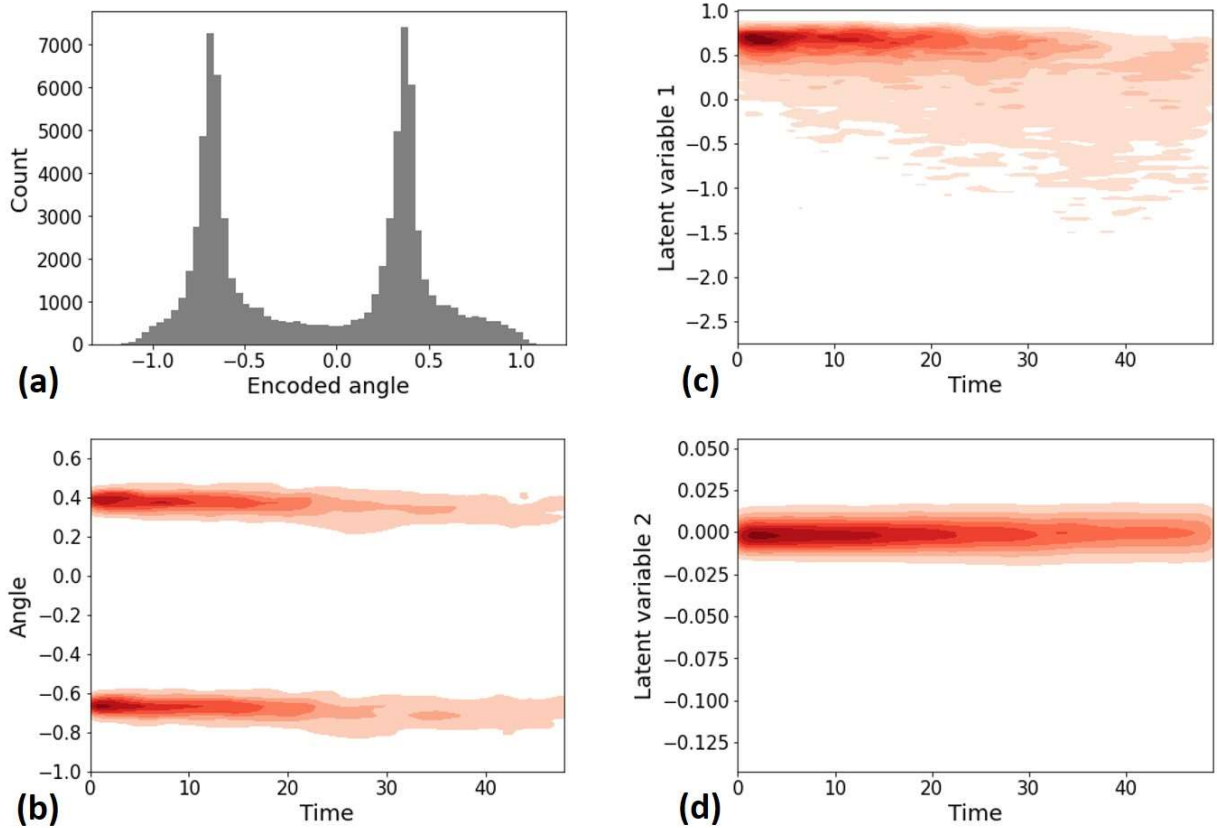

**Figure 7.** Time dynamics of the rVAE components. (a) Average histogram of angle distribution over the full image stack, (b) time-dependent angle distribution, and (c,d) dynamics of 2 latent variables associated with images content.

The time dynamics of the e-beam-induced evolution can be analyzed both in latent space using the time stamp as a parameter and in the time domain. Shown in Fig. 7 (a) is the example of the encoded angle distribution across the whole image stack and (b) its time dependence. Note that the initial sharp bimodal distribution lowers in intensity (due to carbon atom loss) and becomes broader. The evolution of the image latent variables is shown in Fig. 7 (c, d), and demonstrates broadening and reduction in intensity, indicative of lattice decomposition.

**Conclusions**

To summarize, we suggest and implement an approach for the bottom-up description of systems undergoing chemical transformation from atomically resolved imaging data where only partial and uncertain information on atomic positions is available. This approach is predicated on the synergy of two concepts, the parsimony of physical descriptors and the general rotational invariance of non-crystalline solids. The first concept is implemented via the VAE approach, seeking the most effective reduced representation for the system that still contains the maximum original information. The second concept is implementation of rotational invariance. Using this approach, we explore the evolution of e-beam-induced processes in the silicon-graphene system, that can be applied to the studies of formation of small angle grain boundaries, lattice rotation, and amorphization.

We note that this approach can be applied for a much broader range of atomic-scale and mesoscopic phenomena, including domain dynamics, chemical reactivity, and emergence and dynamics of ordered domains where the natural mechanisms of the process necessitates compensation for the continuous rotation of a structural element. The low dimensional latent representations yielded by the autoencoder provide a bottom-up equivalent of the structural order parameters, and hence can be used to describe a broad set of time- and stimulus dependent phenomena in disordered systems in consistent manner. The code used to analyze data is provided as a Jupyter notebook for the interested readers (https://git.io/JfdSa).

**Acknowledgements:**

This effort (ML and STEM) is based upon work supported by the U.S. Department of Energy (DOE), Office of Science, Basic Energy Sciences (BES), Materials Sciences and Engineering Division (S.V.K., O.D., S.J.) and was performed and partially supported (M.Z.) at the Oak Ridge National Laboratory's Center for Nanophase Materials Sciences (CNMS), a U.S. Department of Energy, Office of Science User Facility.

**Authors contribution:**

SVK proposed the concept and led the paper writing. MZ implemented the code used for data analysis and contributed to the paper writing. OD and SJ performed STEM experiments and participated in the interpretation of results.

**Competing Interests:**

All authors declare that they have no competing interests

## Experimental Methods

**Sample preparation:**
Graphene was grown on Cu foil using chemical vapor deposition and spin coated with poly(methyl methacrylate) (PMMA) to preserve the graphene and provide structural stability during handling. The Cu foil was etched away in an ammonium persulfate bath. The remaining PMMA/graphene stack was rinsed, transferred to a TEM grid and the PMMA dissolved in acetone. The sample was baked in a 10% $O^2$ environment at 500 °C for 1.5 hrs to remove contaminant residue.[82] Prior to examination in the STEM the sample was baked again in vacuum at 160 °C for 8 hours.

**STEM imaging:**
STEM imaging was performed using a Nion UltraSTEM 200 operated at 100 kV accelerating voltage with a nominal convergence angle of 30 mrad. The high angle annular dark field detector was used for imaging with a nominal inner angle of 80 mrad. Four holes were drilled through the pristine graphene using by manually positioning the e-beam. An image stack was then acquired capturing the continued evolution of the four holes under the 100 kV e-beam. The beam current was nominally 20 pA and frame acquisition time of ~33.5 s/frame for a dose of $4.2 \times 10^9$ electrons per frame.

**Data analysis:**

*DCNN:*
The DCNN for semantic segmentation of atomically-resolved images had a U-Net architecture[83] and was trained on simulated data using Adam optimizer[84] with the cross-entropy loss function and learning rate of 0.001.

*Convolutional AE:*
In the convolutional AE, the encoder had two convolutional layers with 32 and 64 convolutional filters ("kernels") of size (3, 3) and stride (2, 2) activated by a rectified linear unit (ReLU) function. The decoder consisted of two layers with transposed convolutions of the same size and stride activated by ReLU. The latent layer was represented by a fully connected ("dense") layer with 2 neurons. The model was trained using Adam optimizer with learning of 0.001 and mean-squared error loss function.

*GMM:*
For Gaussian mixture model (GMM), it is assumed that there are $K$ normal distributions parametrized by their mean ($\mu_i$) and covariance ($\Sigma_i$) and that each of extracted sub-images $R_i(x, y)$ is independently drawn from one of these distributions with probability density function given by:

$$p(\boldsymbol{R}_i) = \frac{1}{(2\pi)^{0.5n} |\Sigma_{k_i}|^{0.5}} \exp\left(-\frac{1}{2}(\boldsymbol{R}_i - \mu_k)^T \Sigma_{k_i}^{-1}(\boldsymbol{R}_i - \mu_k)\right).$$

Here, the number of components is fixed (set by an operator) and the expected minimization algorithm is used to determine the parameters of the mixture.

*rVAE:*
The generative and inference models of rotationally-invariant variational autoencoder (rVAE) are parameterized by two deep neural networks, $p_\theta(x|\tau)$ and $q_\phi(\tau|x)$, where $\theta$ and $\phi$ are the parameters (weights and biases) of the neural networks, and $\tau$ denotes a collective latent variable (image content + translation + rotation). A stochastic gradient descent is used to learn the $\theta$ and $\phi$ parameters via maximizing the evidence lower boundary (ELBO) consisting of the mean squared error term and the Kullback-Leibler divergence terms associated with image content and rotation,

$$ELBO = -MSE - D_{KL}(q(z|x)\|\mathcal{N}(0,I)) - D_{KL}(q(\gamma|x)\|\mathcal{N}(0,s_\gamma^2)),$$

where $z$ is a latent variable associated with image content, $\gamma$ is a latent angle variable, and $s_\gamma$ is a "rotational prior". Here, we used encoder and decoder both consisting of 2 fully-connected layers with 128 neurons in each layer activated by *tanh()*. The latent layer had 3 neurons for absorbing arbitrary rotations and translations of images content, whereas the remaining 2 neurons in the latent layer were used for disentangling the image content. The rVAE was trained using the Adam optimizer with the learning rate of 0.0001.

All deep learning routines were implemented in PyTorch deep learning library.[85] For more details, see the interactive Jupyter notebooks reproducing all the data analysis available without restrictions at https://git.io/JfdSa.

**Data availability**:
The experimental data is available without restrictions at 10.5281/zenodo.4279127.